\documentstyle[prd,aps,epsf,floats]{revtex} 
\draft

\begin{document}
\twocolumn[\hsize\textwidth\columnwidth\hsize\csname
@twocolumnfalse\endcsname
\title{
\hbox to\hsize{\large Submitted to Phys.~Rev.~D \hfil E-Print
astro-ph/9704204}
\vskip1.55cm
Maximum Likelihood Analysis of Clusters of
Ultra-High Energy Cosmic Rays}
\author{G\"unter Sigl, Martin Lemoine, and Angela V. Olinto}
\address{Department of Astronomy \& Astrophysics\\
Enrico Fermi Institute, The University of Chicago, Chicago, IL
60637-1433\\}
\date{\today}
\maketitle
\begin{abstract}
We present a numerical code designed to conduct a likelihood
analysis for clusters of nucleons above
$10^{19}\,$eV originating from discrete astrophysical sources
such as powerful radio galaxies, $\gamma-$ray bursts or
topological defects. The code simulates the propagation of
nucleons in a large-scale magnetic field and constructs the
likelihood of a given observed event cluster as a function of the
average time delay due to deflection in the magnetic field, the
source activity time scale, the total fluence of the source,
and the power law index of the particle injection spectrum.
Other parameters such as the coherence length and the 
power spectrum of the
magnetic field are also considered. We apply it to the three
pairs of events above $4\times10^{19}\,$eV recently reported by
the Akeno Giant Air Shower Array (AGASA) experiment, assuming
that these pairs were caused by nucleon primaries which
originated from a common source. Although current data are too
sparse to fully constrain each of the parameters considered,
and/or to discriminate models of the origin of ultra-high energy
cosmic rays, several tendencies are indicated. If the clustering
suggested by AGASA is real, next generation experiments with
their increased exposure should detect more than $\sim10$
particles per source over a few years and our method will put
strong constraints on both the large-scale magnetic field
parameters and the nature of these sources.
\end{abstract}
\vskip2.2pc]


\narrowtext

\section{Introduction}
Despite more than 30 years of experiments on ultra-high energy 
cosmic rays (UHECRs), i.e., air showers with primary
energy $E\gtrsim1\,$EeV ($=10^{18}\,$EeV), the origin of these particles
remains unknown. The most distinctive feature of the energy
spectrum of UHECRs is undoubtedly the ``ankle'',
at $E\simeq10^{18.5}\,$eV~\cite{Bird1,Yoshida1}.
The cosmic ray spectrum is steep up to the ankle, corresponding
to a differential power law index of $\simeq-3.1$, 
and flattens beyond, with an index $\simeq-2.7$.
Data from the Fly's Eye experiment also suggest that the
chemical composition is dominated by heavy nuclei, presumably
iron, up to the ankle, and  by protons beyond~\cite{Bird1}.
Hence, the ankle might mark a transition from galactic to
extra-galactic origin~\cite{Bird1,Yoshida1}.
This idea is supported by the observed isotropy of cosmic rays
for $E\gtrsim4\times10^{19}\,$eV~\cite{Hayashida1},
together with the fact that charged particles with
$E\gtrsim10^{19}\,$eV have too large a gyroradius in the galactic
magnetic field to be confined. However, in analogy to the
$\gamma-$ray burst (GRB) phenomenon, an origin of UHECRs in an
extended galactic halo cannot be excluded thus far.

If UHECRs with
$E\gtrsim10^{18.5}\,$eV are indeed protons of extra-galactic
origin, one would expect to detect the so-called
Greisen-Zatsepin-Kuzmin cutoff~\cite{GZK} (hereafter GZK)
due to photopion production on the cosmic microwave 
background (CMB) by nucleons of energy $E\gtrsim70\,$EeV.
The presence of this cutoff has been suggested by various
experiments, although the energy spectrum around
$E\simeq100\,$EeV is yet to be agreed upon. Nonetheless, the
combined data of the Haverah Park, Yakutsk, Fly's Eye, and AGASA
experiments give 7 detected events above $100\,$EeV, for 22
expected from extrapolation of lower energy data, which provides
a strong hint toward the presence of a cutoff~\cite{Yoshida1}.
Finally, a $2\sigma$ detection of this GZK cutoff has been 
recently claimed by the AGASA experiment from the combined data 
1991-1996~\cite{AGASA-texas}.

Photopion production on the CMB limits the range of nucleons 
above $100\,$EeV to about $30\,$Mpc, whereas heavy nuclei are 
photodisintegrated on an even shorter distance scale~\cite{PSB}.
In this frame, the detection of UHECRs with
$E\gtrsim100\,$EeV~\cite{Bird2,Bird1,Bird3,Hayashida2,Yoshida1}
has triggered considerable discussion in the literature on
the nature and origin of these particles~\cite{SSB,ES,HVSV}.
Indeed, even the most powerful astrophysical objects such as
radio galaxies, quasars, and active galactic nuclei are
barely able to accelerate charged particles to such
energies~\cite{Hillas}. Moreover, there is no obvious optical or
radio counterpart to these UHECR events at a distance less
than a few hundred Mpc. For the highest energy event observed, 
$E\simeq3\times10^{20}\,$eV~\cite{Bird2}, the quasar 3C 
147 is the best candidate as far as energetics is concerned;
however, it lies $\sim1$Gpc away, and if the primary were a 
proton, its injection energy would have to be $E>10^{22}\,$eV. A
similar problem arises for the less likely option of a
$\gamma-$ray primary~\cite{HVSV}, whereas neutrino primaries in
general imply too large a flux because of their small
interaction probability in the atmosphere~\cite{SL}.

Currently, there are three main classes of models for the origin 
of UHECRs. The most conventional one assumes first order Fermi
acceleration of protons at astrophysical magnetized shocks
(see, e.g., Ref.~\cite{BE}). This mechanism is
usually associated with prominent astrophysical objects
such Fanaroff-Riley Class II radio galaxies, either 
in the hot spots or in the lobes~\cite{RB}
and could accelerate protons up to $E\sim10^{21}\,$eV.

It has also been suggested that UHECRs could be associated
with cosmological $\gamma$-ray bursts
(GRBs)~\cite{Waxman1,Waxman2,Vietri1,MU}. This is mainly
motivated by the fact that the required average rate of energy
release in $\gamma-$rays is comparable to the one in
UHECRs above $10\,$EeV turn out to be comparable. Protons could be
accelerated beyond $100\,$EeV within the relativistic shocks
associated with fireball models of cosmological GRBs~\cite{GRB}.
Since the rate of
cosmological GRBs within the field of view of the cosmic ray
experiments which detected events above $100\,$EeV is about 1
per 50 yr, a dispersion in UHECR arrival times of at least
$50\,$yr is necessary to reconcile the observed UHECR and
GRB rates. Such a dispersion could be caused by the time
delay of protons acquired through their deflection in large-scale
magnetic fields (LSMF)~\cite{Waxman1,Waxman2}. However, this
requirement could be circumvented in a model where UHECRs are
produced in a galactic halo population of GRBs~\cite{Vietri2}.

Finally, so-called ``top-down'' models constitute another class
of scenario. There, particles are created at extremely high
energy in the first place by the decay of
some supermassive elementary ``X'' particle associated with
new physics near the grand unification scale~\cite{BHS}.
Such theories predict phase transitions in the early
universe that are expected to create topological
defects such as cosmic strings, domain walls or magnetic
monopoles. Although such defects are topologically stable and
would be present today, they could release X particles due
to physical processes such as collapse or annihilation. Among
the decay products of the X particle are jets of
$10^4-10^5$ hadrons most of which are in the
form of pions that subsequently decay into $\gamma$-rays, electrons, and
neutrinos. Only a few percent of the hadrons are expected to be
nucleons~\cite{Hill}. Thus, typical features of these scenarios are
the predominant release of $\gamma$-rays and neutrinos, and
spectra that are considerably harder than in the case of shock
acceleration. For more details about these models, see,
e.g., Ref.~\cite{Sigl}.

Recently, a possible correlation of a subset of events above
$40\,$EeV among each other and with the supergalactic plane was
reported by the AGASA experiment~\cite{Hayashida1}.
Among 20 events with energy above $50\,$EeV, two pairs
of events with an angular separation of less than $2.5^\circ$
were observed within $10^\circ$ of the supergalactic plane at an
angular resolution of $\simeq1.6^\circ$.
The probability for that to happen by chance for an
isotropic, unclustered distribution is
$\simeq4\times10^{-4}$. A third pair was observed among 36
showers above $40\,$EeV, with a chance probability of about
$6\times10^{-3}$. Although the muon content of the showers 
suggests that the primaries are protons in each case, 
$\gamma$-rays cannot be excluded~\cite{Hayashida1};
in the following, however, we will assume that the primaries are 
protons.

This suggests that the
events within one pair have been emitted by a single discrete
source possibly associated with the large-scale structure of the
galaxies. The deflection of a charged
particle in a magnetic field is inversely proportional to its
energy $E$. Therefore, the fact that the lower energy event
in the pair with the greatest energy difference (see
Table~\ref{pairs_char}) arrived later might hint to the pair's origin
in a burst, i.e., on a time scale $\lesssim1\,$yr.
The time delay would then be dominated by
magnetic deflection of the lower energy
particle. Even the two other pairs observed by AGASA, in which 
the higher energy particle arrived later, could have 
originated in a burst. Since the average time 
delay due to magnetic deflection, $\tau_E$,
scales as $\tau_E\propto E^{-2}$, the dispersion in time 
delay is bounded from below by twice the energy resolution,
$\Delta\tau/\tau \approx$0.6, and could be as
large as $\sim2$~\cite{WM}. Furthermore, the distance
to the source cannot be much larger than $\simeq100\,$Mpc, if the
higher energy primary was either a nucleon, a nucleus, or a
$\gamma$-ray, since its energy was observed to be $\gtrsim75\,$EeV
in all three pairs.

\begin{table}[ht]
\begin{tabular}{lccccc}
  Pair \# & Energy [EeV] & date & $b^{\rm SG}$ & $b^{\rm G}$\\ \hline
  1 & 210 & 93/12/03 & 0.4 & $-41.1$ \\
    & 51  & 95/10/29 & 1.1 & $-42.3$ \\
  2 & 78 & 95/01/26 & 0.5 & 56.9 \\
    & 55  & 92/08/01 & 2.0 & 55.3 \\
  3 & 110 & 94/07/06 & 57.3 & 18.6 \\
    & 43  & 91/04/20 & 57.8 & 21.2 \\
 \end{tabular}
\bigskip
\caption{Properties of the pairs observed by AGASA.
$b^{\rm SG}$ and $b^{\rm G}$ are the supergalactic and the
galactic latitude, respectively.}
\label{pairs_char}
\end{table}

Possible explanations of these observations rest not only on the
model of UHECR origin but also on the largely unknown features
of the LSMF (for a review see, e.g., Ref.~\cite{Kronberg}).
Here, we call LSMF both extra-galactic fields with strengths
$10^{-12}\,{\rm G}\lesssim B_{\rm rms}\lesssim10^{-9}\,$G as
well as fields in the
halo of our Galaxy with $10^{-8}\,{\rm G}\lesssim
B_H\lesssim10^{-6}\,$G. In
the present paper, we show how the distribution of arrival times
and energies of clusters of UHECR events
originating from the same astrophysical source such as the ones
suggested by AGASA can be used to constrain both the nature of
the source and the structure of the LSMF. We restrict ourselves
to nucleons and perform Monte Carlo simulations of their
propagation through the LSMF. Then we construct the
likelihood as a function of several parameters characterizing
the LSMF and the source of the pairs observed by AGASA.
We describe our Monte-Carlo simulations and the likelihood
analysis in Sect.~2. In Sect.~3, we discuss our findings and
their implications for the structure of the LSMF and for the
models of UHECR origin. Sect.~4 concludes with a discussion of
the future prospects of the method presented in this
paper.

\section{Numerical simulations}
\subsection{The Monte-Carlo code}

Previous works have focused on various aspects of non-diffusive,
nearly straight-line propagation of extremely high energy
nucleons in LSMF. In Ref.~\cite{WM} the diffusion problem was
studied, in the limits $D\theta_E\ll
l_{\rm c}$ and $D\theta_E\gg l_{\rm c}$, where $D$ is the distance to the 
source, $\theta_E$ is the r.m.s. deflection angle, which is a
function of energy $E$, and $l_{\rm c}$ is the coherence length of the
magnetic field. However, these authors did not consider pion 
production, and their results are only valid for energies 
$E\lesssim50\,$EeV. In Refs.~\cite{YT,PJ,Lee} propagation codes
that follow multiple species were constructed, taking all
types of interactions into account, although 
neglecting any scattering on the magnetic field; thus these
calculations cannot treat the time delay of UHECRs. Finally, 
the authors of Ref.~\cite{TPH} carried out tri-dimensional 
Monte-Carlo simulations, including pion production effects and 
scattering off magnetic field inhomogeneities. In their work,
pion production is treated as a continuous energy loss process
and its stochastic nature is neglected. The magnetic 
field configuration is simplified to an assembly of coexistent 
bubbles of randomly aligned dipole fields, with a diameter
given by the coherence length of the field.

To properly calculate the desired likelihood, one has to
consider both pion production and
deflection on cosmic magnetic fields. Thus we devised
a Monte Carlo code, which improves on the treatment in
Ref.~\cite{TPH} on the points discussed in the following.

Pion production is treated in the
appropriate stochastic way, as described in Ref.~\cite{Lee};
namely, the occurrence, and the nature of the secondary
and its energy are randomly drawn
from the relevant probability distributions. Since the
fractional energy loss of a proton in a pair production event is
about the ratio of the electron to the proton mass, pair 
production can be
included into the equations of motion as a continuous energy
loss term, for which we used the expressions in Ref.~\cite{CZS}.

The LSMF are described as gaussian random fields,
with  zero ensemble average, and a power spectrum given by
$\left\langle B^2(k)\right\rangle\propto k^{n_B}$ for
$k<2\pi/l_c$, and $\left\langle B^2(k)\right\rangle=0$ otherwise.
The cutoff, $l_c$, characterizes the
coherence length of the field. The phase and direction of
the field are drawn from uniform random
distributions. Note that for the time delay calculation only the
field component perpendicular to the line-of-sight contributes.  
This model can thus be characterized by three
parameters: the power-law index $n_B$, the coherence scale $l_c$,
and the r.m.s. field  strength, given by
$B_{\rm rms}^2\equiv V/(2\pi)^3\int d^3{\bf k}B^2({\bf k})$.
 
The precise structure of the LSMF is unknown, but
estimates can be made for different cases. Extra-galactic
fields with cosmological origins have
$n_B \simeq 0$ and $l_c\simeq 1\,$Mpc~\cite{JKO}.
For halo fields to cause notable delays, $l_c\gtrsim1\,$kpc. 
Note that for these large coherence lengths the
dynamical range probed by UHECRs is not very sensitive to $n_B$.

In addition to the LSMF defined above, there are localized fields
associated with galaxies, clusters of galaxies, and the large
scale galaxy distribution.  The small filling factors of
galaxies, $f_G\sim10^{-4}$, and clusters, $f_C\sim10^{-6}$,
justify neglecting their effect. If galactic and cluster outflows
are efficient polluters of the extra-galactic medium, their effect
can be modeled by the extra-galactic field considered above. 
The filling factor for such pollution fields was estimated to
$\simeq10$\% for a field strength considerably smaller than 
$\sim\mu{\rm G}$~\cite{KL}.

Fields associated with the galaxies large scale structure may
significantly affect UHECR propagation. In fact, the arrival
directions of two of the three pairs observed by AGASA (see
Table~\ref{pairs_char})
are very close to the supergalactic plane; reinforcing the claim
for a supergalactic plane correlation of UHECRs~\cite{Stanevetal}.
As was subsequently pointed out in
Ref.~\cite{WFP}, this correlation seems to
be even stronger than expected if the origin of UHECRs 
were associated with the large-scale galaxy structure, as
the supergalactic plane alone is not a good approximation of 
this latter. As a solution, it has been suggested, in
Refs.~\cite{BKR,Rachen},
that the possible existence of strong fields, with strengths up
to $\mu$G and coherence lengths in the Mpc
range, aligned along the large-scale structure,
could produce a focusing effect of UHECRs along the sheets and
filaments of galaxies. However, since UHECR propagation would
take place mainly along the strong coherent field component,
effects on the
time delay would be much smaller than those induced by an
incoherent field of comparable strength. Within a first
approximation, therefore, we can neglect such coherent strong
field components in our present analysis, which rests primarily 
on the time delay effect. In other words, our
analysis is sensitive mainly to the weaker random field components
that are not aligned with the large-scale structure sheets and
filaments.

We also neglect the influence of the magnetic field
in the disk of our Galaxy.
Denoting its typical strength by $B_{\rm d}$ and its scale by
$l_{\rm d}$, the deflection angle $\theta_E$ for a proton at
energy $E$ arriving from above or below the galactic disk is
\begin{equation}
  \theta_E\simeq5.5^\circ
  \left({E\over10\,{\rm EeV}}\right)^{-1}
  \left({l_{\rm d}\over1\,{\rm kpc}}\right)
  \left({B_{\rm d}\over10^{-6}\,{\rm G}}\right)\sin\theta\,,\label{gal1}
\end{equation}
where $\theta$ is the angle between the field polarization and
the arrival direction of the particle. The corresponding time
delay, $\tau_E\simeq\theta_E^2l_{\rm d}/2$, is small compared to the
measurement period of a few years and can be neglected for
$E\gtrsim10\,$EeV in our analysis, if $l_{\rm g}\lesssim100\,$pc
and $B_{\rm d}\lesssim10^{-6}\,$G, which are typical values for the
disk field. We also remark that none of the AGASA pair arrival
directions lie close to the galactic plane (see
Table~\ref{pairs_char}). We will assume that
the actual time delay is dominated either by the extra-galactic
magnetic field or by an extended galactic halo field. Under this
assumption, observed time dispersions contain information only
about the extra-galactic magnetic field or the galactic halo
field~\cite{LSOS}.

Once the configuration of the LSMF is set, via Fast Fourier 
Transform methods, on a lattice with inter-cell separation
$l_{\rm c}/2$ and typical size $64^3$ or $128^3$, a sample of 
nucleons is propagated over a distance $D$ in a randomly chosen
direction. The integration is achieved via the 
usual fourth order adaptive step size Runge-Kutta integration 
(see, e.g., Ref.~\cite{Press}).
At each time step, the nucleons are subjected to 
deflection by the local magnetic field, are tested against pion 
production and $\beta$ decay if the nucleon is a neutron,
and pair production energy loss is taken into account.

\begin{figure}
\centering\leavevmode
\epsfxsize=3.2in
\epsfbox{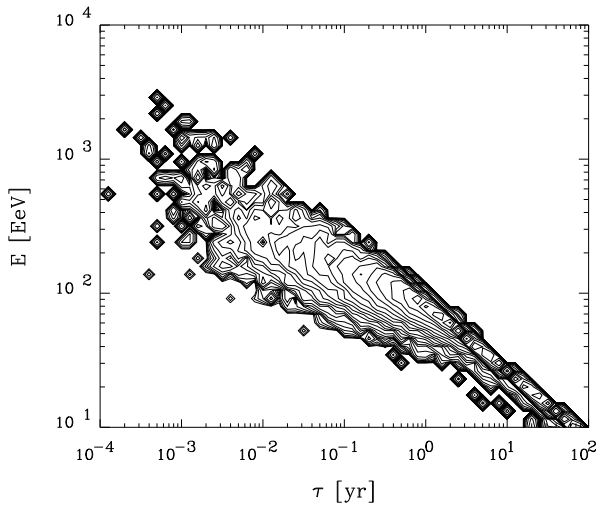}
\caption{Contour plot for a distribution of events originating 
from a bursting source, as projected in the time-energy plane. 
The corresponding parameters are $D=30\,$Mpc, 
$B_{\rm rms}=10^{-11}\,$G, $n_B=0$, 
$l_{\rm c}=1\,$Mpc; the differential index of the energy 
injection spectrum is $-2.0$. Deviations from the mean 
correlation $\tau_{\rm E}\propto E^{-2}$,
for $E\geq70\,$EeV,
reflect pion production on the CMB. In order to show the 
statistics at high energies, 40 contours with logarithmic 
decrements of 0.15 in the logarithm to base 10, are
shown.}
\label{F1}
\end{figure}

Angle-time-energy images of a discrete source at a fixed
location depend on the specific LSMF structure which the
nucleons encounter during their propagation~\cite{LSOS}. 
In principle, one
has to simulate a discrete source by emitting particles
isotropically from the origin, and keeping those particles
that arrive on the sphere of radius $D$ within a solid angle
segment small compared to the squared r.m.s. deflection angle.
This is crucial in the limit of small
deflection, $D\theta_E\ll l_{\rm c}$; this limit is also the 
most probable for typical extra-galactic magnetic fields
(see Eq.~(\ref{t_delay}) below, see also 
Ref.\cite{LSOS}). In the opposite limit,
$D\theta_E\gg l_{\rm c}$, where nucleons following different 
paths experience different magnetic field configurations, it is 
sufficient to send the particles isotropically, 
and average over all events independently of their arrival 
location. The effect of these two limits on the resulting 
angle-time-energy image of the source was pointed out in 
Ref.\cite{WM}. In the limit of small deflection, the scatter 
around the mean correlations, $\tau_E\propto E^{-2}$ and 
$\theta_E\propto E^{-1}$, is expected to be negligible below the 
photopion production threshold, with $\Delta E/E\sim 1$\%. In 
the opposite limit, $D\theta_E\gg l_{\rm c}$, the stochastic
deflections of different nucleons off different magnetic 
inhomogeneities imply a significant scatter, with 
$\Delta E/E\sim30$\%. We confirmed that our code yields
ditributions in energy and time that are consistent with the
analytical expression given in Ref.~\cite{WM} for this case,
below the GZK cutoff.

In order to simulate the small deflection limit, one has to
restrain oneself to defining a receiving cone angle 
smaller than the r.m.s. deflection angle, and to selecting an
emission cone angle not much larger than the receiving angle.
This will in general introduce a bias in the simulated energy
spectrum, notably because the r.m.s. deflection angle is a
function of energy. In order to correct for this bias, we 
renormalize the time-integrated flux at a given
energy to the corresponding flux for vanishing magnetic fields.
This is appropriate in our case, where delay times are much
smaller than energy loss distances, i.e., where the 
magnetic field has no influence on the time-integrated energy 
spectrum. Fig.~\ref{F1} shows an example for the resulting
distributions of arriving particles in time and energy for a 
bursting source, in the small deflection limit. For a 
more general discussion of the angle-time-energy images 
of UHECR sources, see Ref.~\cite{LSOS}.

\subsection{Likelihood}

We wish to evaluate the likelihood of the observations of the 
three AGASA pairs as a function of the different parameters 
characterizing the origin of UHECRs and the nature of their 
energy spectrum, as well as those characterizing the LSMF.
We thus consider the following. We assume that the source 
emits UHECRs on the time scale $T_{\rm S}$; for a burst, 
$T_{\rm S}\ll1$yr. We group together the distance to 
the source, $D$, the coherence length $l_{\rm c}$, the r.m.s.
magnetic field strength $B_{\rm rms}$, and the magnetic power
spectrum index $n_B$, in the average time delay $\tau_E$; this 
latter is calculated from the expressions given in 
Ref.~\cite{WM}, as:
\begin{eqnarray}
  \tau_E\,&\simeq&\,
  0.22\,\left(\frac{3+n_B}{2+n_B}\right)
  \left(\frac{D}{30\,{\rm Mpc}}\right)^2
  \left(\frac{E}{100\,{\rm EeV}}\right)^{-2}\nonumber\\
  &&\times\left(\frac{B_{\rm rms}}{10^{-11}\,{\rm G}}\right)^2
  \left(\frac{l_{\rm c}}{1\,{\rm Mpc}}\right)\;{\rm yr}.
  \label{t_delay}
\end{eqnarray}
This follows from a random walk argument, and is, strictly 
speaking, a good approximation only below the pion production
threshold and in the large deflection limit,
$D\theta_E\gg l_{\rm c}$.
We also consider the distance $D$ as a separate parameter, as it
governs the amplitude of pion production, hence the propagated
UHECR energy spectrum.

For the total energy output of the source in UHECRs
above an energy $E$, we assume a power law,
${\cal E}(>E)\propto E^{-\gamma+2}$, and consider three
different representative values for the index $\gamma$, namely
$\gamma=1.5,\,2.0,$ and $2.5$.
A hard spectrum with $\gamma=1.5$ is typical for defect
models~\cite{BHS,Sigl}, whereas softer spectra with 
$\gamma\gtrsim2.0$ are typical for acceleration at astrophysical 
shocks~\cite{BE}. With respect to observations, it is more
convenient to characterize the total energy output with the total
number of particles $N(E_0)$, that would be detected by the
experiment with an energy $E\geq E_0$, over an infinite
integration time,
\begin{eqnarray}
  N(E_0)&\simeq&2\times10^2\,f_{\rm d}\left(\frac{A}{100\,{\rm km}^2}\right)
  \left(\frac{D}{30\,{\rm Mpc}}\right)^{-2}\label{fluence}\\
  &&\times f_{\rm s}(E_0)\left(\frac{E_0}{10^{1.5}\,{\rm EeV}}\right)^{-1}
  \left(\frac{{\cal E}(>E_0)}{10^{51}\,{\rm erg}}\right)
  \,.\nonumber
\end{eqnarray}
Here, $f_{\rm d}$ denotes the experimental duty cycle with respect to 
having the source in the field of view, $A$ is the effective
detection area of the experiment under consideration, and
$f_{\rm s}(E_0)$ is a dimensionless factor of order unity 
that depends weakly on the energy injection spectrum cutoff.
Eq.~(\ref{fluence}) is valid
below the pion production threshold and in the following we
take $E_0=10^{1.5}\,$EeV which is the energy above which the 3
AGASA pairs have been observed, within the energy resolution. We
consider values for $N(10^{1.5}\,{\rm EeV})$ between 1 and
$4000$ particles, which is a typical range to be expected for
the fiducial values in Eq.~(\ref{fluence}).

The main parameters defining the likelihood are therefore
$T_{\rm S}$, $\tau_{100}\equiv\tau_{E=100\,{\rm EeV}}$,
$D$, $\gamma$, and $N_0\equiv N(E_0=10^{1.5}\,{\rm EeV})$; secondary
parameters are $l_{\rm c}$ and $n_B$.
The likelihood is calculated in the standard way for each
observed event cluster denoted by $i_{\rm P}$, using Poisson
statistics,
\begin{equation}
  {\cal L}\left(\tau_{100},T_{\rm S},D,\gamma,
  N_0,i_{\rm P}\right)\equiv\prod_{j=1,N}
  e^{-\rho_j}\frac{\rho_j^{n(j,i_{\rm P})}}{n(j,i_{\rm P}){\rm !}}\,,
  \label{likelihood}
\end{equation}
where $\rho_j$ is the predicted number of events in cell $j$, 
and $n(j,i_{\rm P})$ is the number of observed events (either 0 
or 1 in our case) in cell $j$ for the observed event cluster $i_{\rm P}$. 
Each cell is defined by a time coordinate and an 
energy. We binned the time-energy histogram to logarithmic
energy bins of size 0.1 in the logarithm to base 10,
and to $0.1\,$yr in linear time bins (we 
are limited to $0.1\,$yr time resolution because of memory size).
The product in the formula above extends over all energy bins
(from $10^{1.5}\,$EeV to $10^4\,$EeV) and over all time bins that are 
comprised within an observational time window of 
length $T_{\rm obs}$; we took $T_{\rm obs}\simeq5\,$yr, as AGASA started
operating in 1990, and the latest data reported date back
to the end of October 1995.

The likelihood as given above has already been marginalized over 
two auxiliary parameters: a zero-point in time $t_0$, which 
defines the position of the observational time window on the 
time delay histogram of the UHECRs, and the path that the 
particle followed through the magnetic field. The first 
marginalization over $t_0$ is carried out by drawing $t_0$
at random a large number of times from a uniform distribution,
calculating the likelihood, and averaging the resulting 
likelihood over $t_0$ with equal weights. The second 
marginalization is carried out in a similar spirit, i.e., 
by separately calculating likelihoods for different realizations
of the LSMF between the source and the
observer and averaging them with equal weights.

The likelihood defined in Eq.~(\ref{likelihood}) is thus 
calculated in the following way: for each 
parameter (except $T_{\rm S}$, $\gamma$, and $N_0$: see below), and
before each marginalization, typically $2\times10^4$ particles
are propagated. The resulting histogram in time and energy is 
smeared out in energy with the AGASA energy resolution
$\Delta E/E\sim30$\%; this histogram is also convolved on the 
time axis with a top-hat of width $T_{\rm S}$ to represent a uniform 
emission of particles during the time $T_{\rm S}$. With the same 
population of propagated nucleons, one can construct different 
histograms depending on $\gamma$ and $N_0$. The nucleons are 
indeed injected with an artificially flat spectrum between
$10^{1.5}\,$EeV and $10^4\,$EeV, corresponding
to $\gamma=0$ to preserve satisfying statistics at high 
energies, and the counts on the detector are weighted in such a 
way as to reconstruct an injection spectrum with index $\gamma$.
The likelihood is then marginalized over $t_0$, and calculated 
for each $T_{\rm S}$. New samples of nucleons are propagated to 
marginalize the likelihood over the different magnetic field
realizations; finally, the last loop closes, the above process 
is repeated for each value of $\tau_{100}$, and
$D$. Due to the multi-dimensional nature of this likelihood,  
the size of the parameter space, notably the possible distances
to the source, and the precision required ($1\,$yr corresponds to 
$3\times10^{-7}\,$Mpc), these simulations are very time 
consuming. They were carried out on IBM at the Max-Planck
Institut f\"ur Physik, M\"unchen (Germany), and on
Alpha at the Institut d'Astrophysique de Paris, Paris 
(France).

We probe values for $\tau_{100}$ between $\simeq0.1\,$yr and
$1000\,$yr. The lower bound is given by the size of our time bins
whereas the upper bound is motivated by Faraday
rotation limits on the extra-galactic magnetic field (e.g.,
Ref.~\cite{Kronberg}), which can be written as
$(B_{\rm rms}/10^{-9}\,{\rm G})(l_{\rm c}/1\,{\rm Mpc})^{1/2}
\lesssim1$, or,
\begin{equation}
  \tau_{100}\lesssim2.5\times10^3\,\left(\frac{D}{30\,{\rm
  Mpc}}\right)^2\,{\rm yr}\label{Faraday}
\end{equation}
[see Eq.~(\ref{t_delay})]. A comparable bound comes from the
requirement that the relative deflection of the events in the
AGASA pairs was smaller than the angular resolution
$\simeq1.6^\circ$~\cite{Cronin2}.

As to the source activity
time scale, we choose the range $0.01\,{\rm yr}\leq
T_{\rm S}\leq10^3\,$yr which, together with $1\leq N_0\leq4000$,
brackets the range where the likelihood function should peak,
given a couple of events observed within $5\,$yr.
We note that, as long as the total UHECR output of the source,
characterized by ${\cal E}(>10^{1.5}{\rm EeV})$ or
$N_0$ [see Eq.~(\ref{fluence})], is unconstrained, the
likelihood analysis cannot distinguish between different values
for $T_{\rm S}$ and $\tau_{100}$ that are considerably larger
than the integration time $T_{\rm obs}$ of the experiment.
In this limit, the likelihood becomes degenerate and only
depends on the ratio(s)
$N_0/T_{\rm S}$ or $N_0/\tau_{100}$, i.e., the actual 
fluxes as seen by the experiment. Strictly speaking, it would 
be more appropriate to introduce the deviation around 
$\tau_{100}$ in the latter ratio, instead of $\tau_{100}$; 
however, at a given energy, this deviation is always a finite 
fraction of $\tau_{100}$ whose size only depends on whether the 
deflection is large or small compared to the coherence 
length~\cite{WM}.

It is interesting to note that pairs 2 and 3 do not show 
arrival energies inversely correlated to their arrival times, 
as would be expected from Eq.~(\ref{t_delay}), whereas pair 1 
does. As mentioned in the introduction, this flip around of time 
and energy for these pairs could be explained by the finite 
width of the $\tau_E-E$ correlation, i.e., the combination
of the scatter due to the stochastic nature of the magnetic
field, and of the finite instrumental energy resolution. 
Nevertheless, this could still be insufficient for pairs 2 and
3; in this case, these 
pairs would tend to favor a continuous source, where a flip 
around can simply result from different emission times. We will 
observe each of these tendencies in Sec.4.

Finally, we restrict ourselves to the limit of small deflections,
$D\theta_E\ll l_{\rm c}$. In effect, since the three pairs
observed by AGASA imply $\theta_E\lesssim2.5^\circ\simeq0.044$, 
the limit $D\theta_E\gg l_{\rm c}$ would require
$D/l_{\rm c}\gg25$ and thus a rather large lattice in our
code. Moreover, since the AGASA pairs also imply
$D\lesssim50\,$Mpc, due to pion production,
this case would only be realized if $l_{\rm c}\ll2\,$Mpc. This 
is not likely in the case of an extra-galactic magnetic field, 
due to possible damping of the field on small scales 
$\lesssim1\,$Mpc~\cite{JKO}. If the deflection of UHECRs is
dominated by the halo magnetic field, the distance to the source 
in Eq.~(\ref{t_delay}) has to be replaced by the linear
extension of the magnetic halo $l_{\rm H}$
(if the source lies outside of this halo); in 
this case, the limit $D\theta_E\gg l_{\rm c}$, would necessitate
$l_{\rm c}\ll4\,$kpc if $l_{\rm H}\lesssim100\,$kpc.

\section{Results}

\begin{figure}
\centering\leavevmode
\epsfxsize=3.2in
\epsfbox{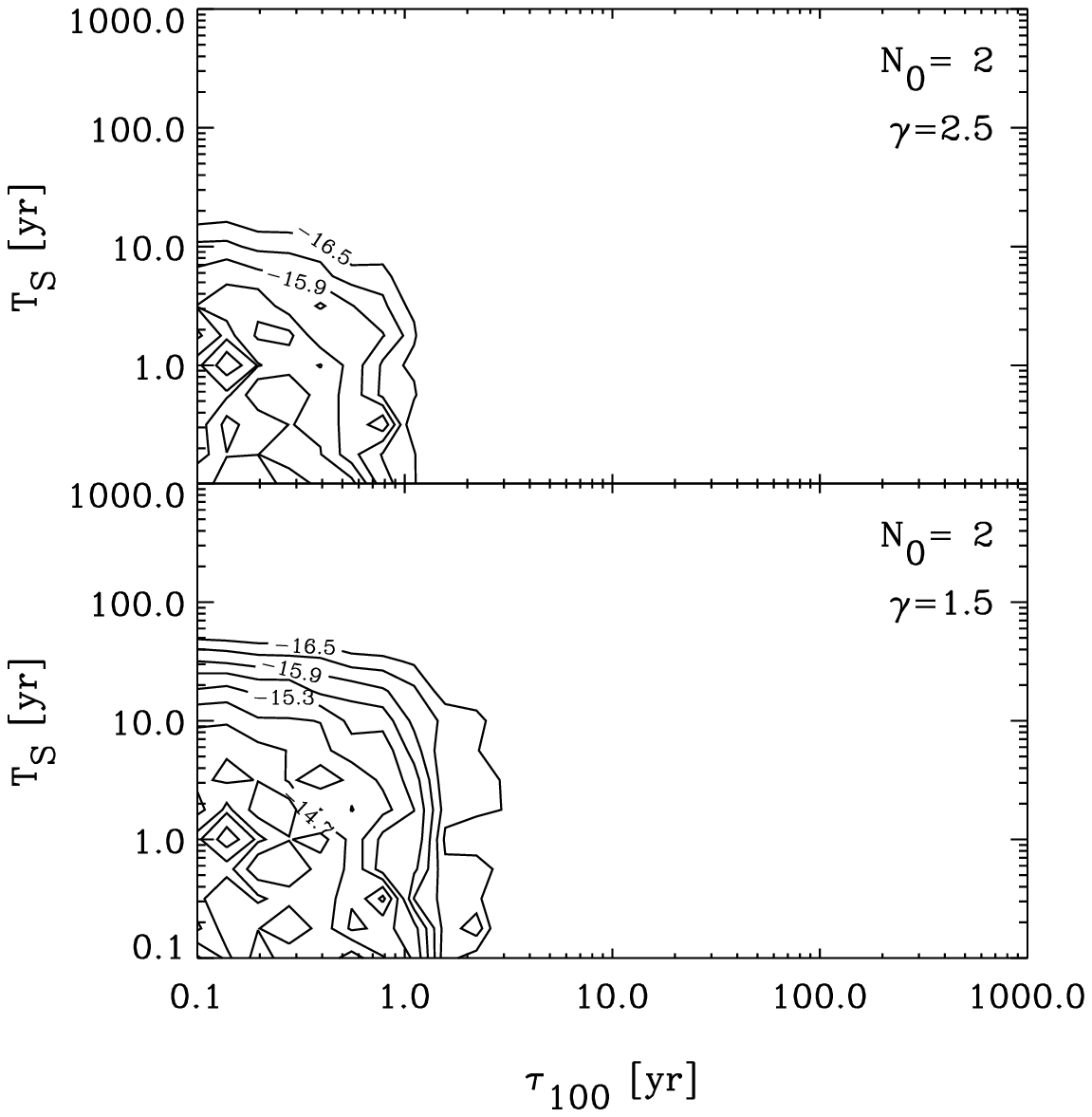}
\caption{Contour plots of the log-likelihood  
$\log_{10}{\cal L}$ in the $\tau_{100}-T_{\rm S}$
plane for pair 1. The corresponding parameters are 
$D=30\,$Mpc, $n_B=0$, $l_{\rm c}=1\,$Mpc; $N_0$ and $\gamma$ 
as indicated. The bottom panel corresponds to the maximum of the 
likelihood.}
\label{F2}
\end{figure}

\begin{figure}
\centering\leavevmode
\epsfxsize=3.2in
\epsfbox{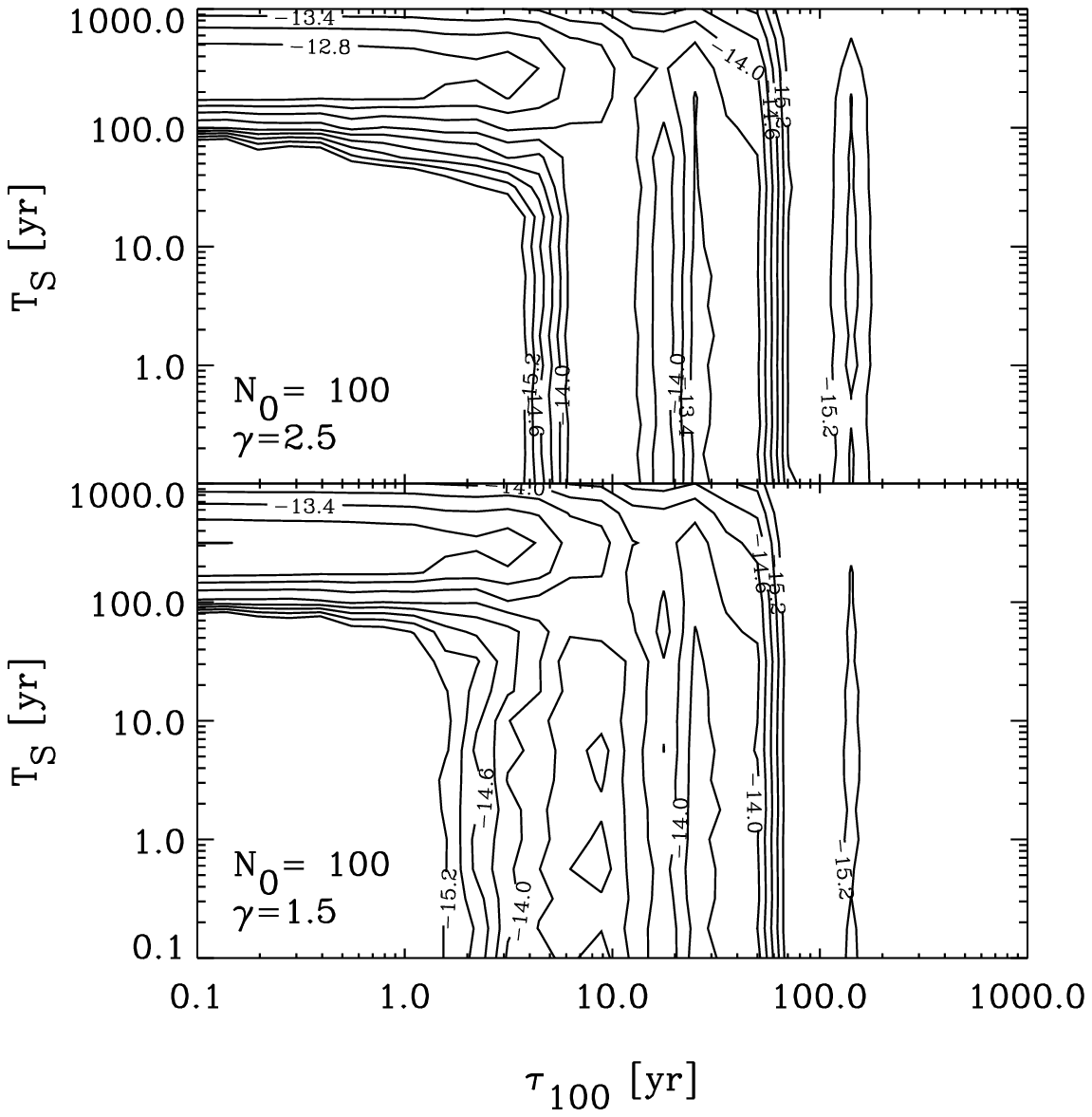}
\caption{Same as Fig.~2, but for pair~2.}
\label{F3}
\end{figure}

\begin{figure}
\centering\leavevmode
\epsfxsize=3.2in
\epsfbox{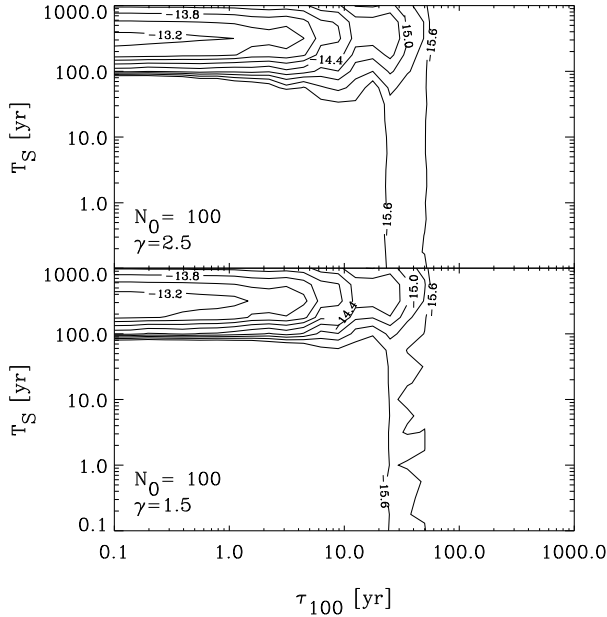}
\caption{Same as Fig.~2, but for pair~3.}
\label{F4}
\end{figure}

\begin{figure}
\centering\leavevmode
\epsfxsize=3.2in
\epsfbox{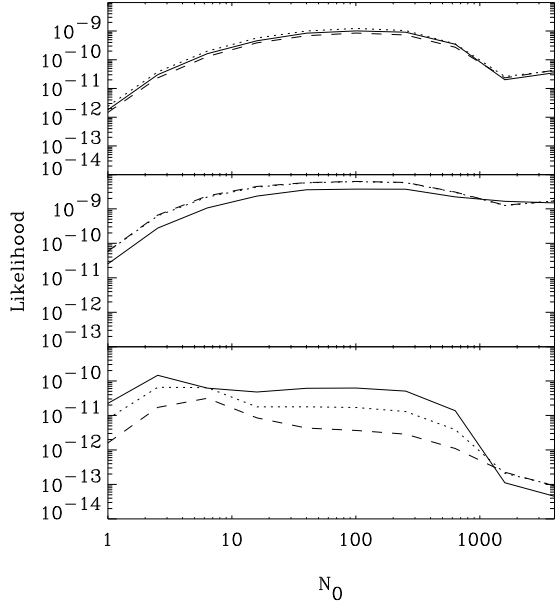}
\smallskip
\caption{Likelihood marginalized with respect to $\tau_{100}$ and
$T_{\rm S}$, plotted {\it vs} $N_0$, for pair 1 (bottom 
panel), pair 2 (middle panel), and pair 3 (top panel). The solid 
lines correspond to $\gamma=1.5$, the dotted lines to 
$\gamma=2.0$, and the dashed lines to $\gamma=2.5$. The
parameters $D$, $n_B$, and $l_{\rm c}$ are the same as in
Figs.~2$-$4. The decrease for $N_0>500$ for pair~1 and~3 is explained by
the fact that in this range one would expect more than 2
events per $5\,$yr since $T_{\rm S}\leq10^3\,$yr in our
simulations. Due to the lower energies this can be compensated
by a large time delay In case of pair~2.}
\label{F5}
\end{figure}

\begin{figure}
\centering\leavevmode
\epsfxsize=3.2in
\epsfbox{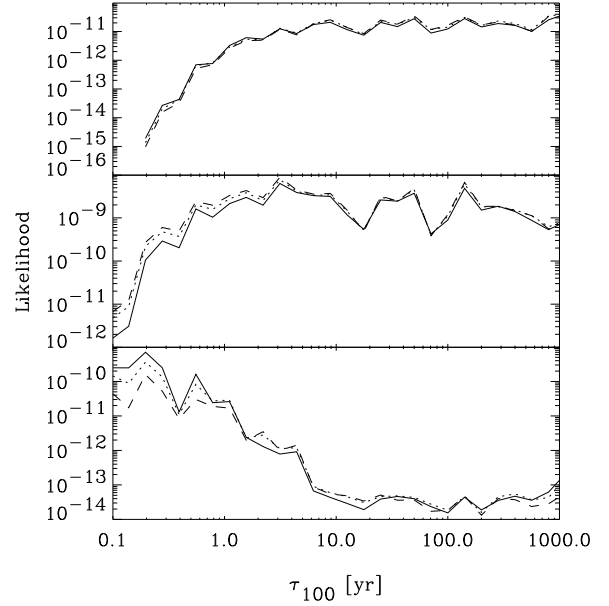}
\smallskip
\caption{Likelihood marginalized with respect to $N_0$,
assuming a bursting source,
$T_{\rm S}\ll 1.\,$yr, plotted {\it vs} $\tau_{100}$, for pair 1
(bottom panel), pair 2 (middle panel), and pair 3 (top panel).
The solid lines correspond to $\gamma=1.5$, the dotted lines to 
$\gamma=2.0$, and the dashed lines to $\gamma=2.5$. The
parameters $D$, $n_B$, and $l_{\rm c}$ are the same as in
Figs.~2$-$4.
This particular marginalization allows to estimate the most
probable time delay, assuming a bursting source.}
\label{F6}
\end{figure}

\begin{figure}
\centering\leavevmode
\epsfxsize=3.2in
\epsfbox{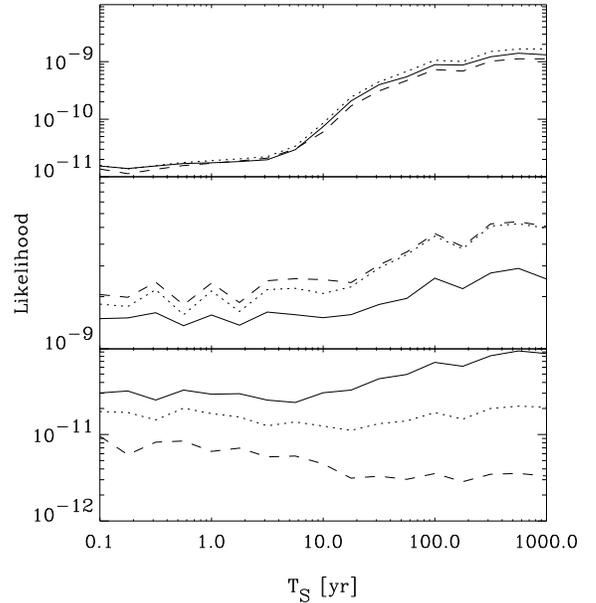}
\smallskip
\caption{Likelihood marginalized with respect to $\tau_{100}$ 
and $N_0$, plotted {\it vs} $T_{\rm S}$, for pair 1 (bottom 
panel), pair 2 (middle panel), and pair 3 (top panel). The solid 
lines correspond to $\gamma=1.5$, the dotted lines to 
$\gamma=2.0$, and the dashed lines to $\gamma=2.5$. The
parameters $D$, $n_B$, and $l_{\rm c}$ are the same as in
Figs.~2$-$4.}
\label{F7}
\end{figure}

Since the likelihood function depends on so many variables we
will present several one- and two-dimensional cuts through the
parameter space. Because the values of most of the parameters
are basically unconstrained it is also convenient to marginalize
over one or several of the variables. In order to do so, one
has to choose a suitable Bayesian prior. We will always assume a
Jeffrey prior, i.e., uniform distributions in
the logarithm
of the variables $\tau_{100}$, $T_{\rm S}$, and $N_0$. This 
choice is motivated by the ``scale invariance'' of the problem, 
namely the fact that the time delay depends on powers of the 
magnetic field strength and coherence length, which are totally 
unknown, as well as the fact that the emission time scale and energy 
output are unknown within many orders of magnitude. We will thus
consider the following types of representations in all of which
$D=30{\rm Mpc}$, $n_B=0$, and $l_{\rm c}=1{\rm Mpc}$ take specific
constant values, for illustration:

\begin{enumerate}
\item For a given $N_0$, contour plots of the likelihood
$\log_{10}({\cal L})$ are shown in the
$\tau_{100}-T_{\rm S}$ plane, for two extreme values of the
power-law index of the injection spectrum,
$\gamma=1.5$, and $\gamma=2.5$, see
Figs.~(\ref{F2}$-$\ref{F4}).

\item The likelihood ${\cal L}(N_0)$, as
marginalized over $T_{\rm S}$ and $\tau_{100}$, is shown in 
Fig.~(\ref{F5}).
Via Eq.~(\ref{fluence}), this determines the most likely value
for the total energy output ${\cal E}(>10{\rm EeV})$ of the
source.

\item The likelihood
${\cal L}(\tau_{100},T_{\rm S}\ll0.1{\rm yr})$
marginalized over $N_0$ is shown in 
Fig.~(\ref{F6}). This particular 
marginalization assumes a bursting source, and thus allows to 
determine the most probable corresponding time delay. Notably, 
the model of origin of UHECRs in cosmological $\gamma$-ray 
bursts requires $\tau_{100}>50{\rm yr}$\cite{Waxman1,Waxman2}.

\item The likelihood ${\cal L}(T_{\rm S})$, as
marginalized over $\tau_{100}$ and $N_0$, is shown in
Fig.~(\ref{F7}). This marginalization 
allows to test the respective significance of a bursting or 
a continuous source.

\end{enumerate}

\begin{figure}
\centering\leavevmode
\epsfxsize=3.2in
\epsfbox{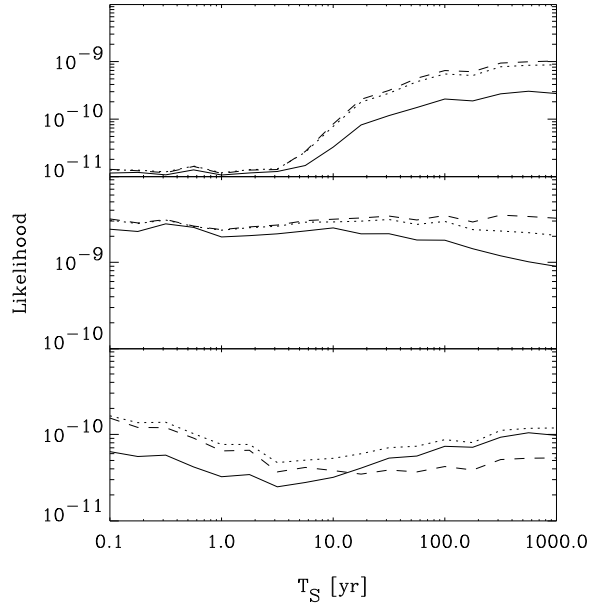}
\smallskip
\caption{Same as Fig.~7, but for a source at $D=100\,$kpc and a
galactic halo field with $n_B=4$, and $l_{\rm c}=30\,$kpc,
extending over $100\,$kpc.}
\label{F8}
\end{figure}

\mbox{ }From the figures shown here it is obvious that there are
too many parameters for the current number of observations
available. However, the following trends are discernible:

\begin{itemize}
\item In the limit of a high number of particles, the typical
time dispersion, i.e., either $T_{\rm S}$ or the deviation
around $\tau_{100}$, scales like $N_0$, so that the particles
are spread out in time on the detector; i.e., so that the 
observed flux matches the injected flux. The likelihood thus
becomes degenerate, and depends only on the ratios
$N_0/T_{\rm S}$, and/or $N_0/\tau_{100}$, as argued in 
Sec.2. This degeneracy arises whenever
$T_{\rm S}\gg T_{\rm obs}$, or the dispersion of the time delay 
for the lower energy particle in the pair, is much larger than 
$T_{\rm obs}$. This can be seen best in
Figs.~\ref{F6}$-$\ref{F8}. In fact, pairs~2 and~3 (see
Table~\ref{pairs_char}), in which the higher energy particle
arrived later, favor, respectively marginally and clearly, a
comparatively large $T_{\rm S}$.

\item Pair~1 marginally (i.e., within factors of a few in the
likelihood) favors a hard injection spectrum,
which is expected in order to produce a $200\,$EeV
event. For $\tau_{100}\lesssim$ a few years
it allows for a bursting source. This is mainly due to the fact 
that its energies and arrival times are inversely correlated, as 
expected from Eq.~(\ref{t_delay}). The small magnitude of the 
corresponding $\tau_{100}\lesssim1\,$yr results from the small 
difference in arrival times as compared to the significant
difference in arrival energies (see
Fig.~\ref{F6}). The origin of this 
pair is therefore not straightforward to account for.

\item Pair~2 is insensitive to $\gamma$, since the energies of
the particles are very close to each other. It allows for a
burst with a large time delay, although the arrival times and 
energies are not inversely correlated. As argued previously, the 
flip around of the energies can be explained by the finite 
instrumental energy resolution. In our case, the instrumental 
resolution $\Delta E/E\sim30$\% dominates over the intrinsic 
scatter~\cite{WM} due to the scattering off 
the magnetic field. This width introduces a dispersion in 
time delay, $\Delta\tau_E/\tau_E=2\Delta E/E$. Hence, 
provided the time delay is sufficiently large, there is 
significant scatter both in energies and arrival times. This 
explains the tendency of pair 2 to accept a large time delay.
Although the marginalized likelihood ${\cal L}(T_{\rm S})$
slightly favors $T_{\rm S}\gtrsim10\,$yr, this is not very
significant and still consistent with a burst (see
Fig.~\ref{F7}). This pair is thus consistent with an origin in
cosmological $\gamma$-ray bursts.

\item Pair~3, however, appears to be completely inconsistent
with a burst for time delays $\tau_{100}\lesssim100\,$yr.
For larger time delays and correspondingly high total expected
number of particles $N_0$, this pair could be consistent
with a burst. However, such high time delays might in turn be
inconsistent with the Faraday rotation limit, as given by 
Eq.~(\ref{Faraday}). Furthermore, the likelihood is suppressed
by roughly a factor of 100 for such a scenario (see
Figs.~\ref{F4} and~\ref{F7}).
Overall, and within the parameter space probed here, 
this pair seems to favor an origin in a continuous source with a 
high total energy output, corresponding to $N_0\gtrsim100$.

\item In all cases the likelihood function tends to peak in the
range where $\tau_{100}\lesssim10\,$yr. This is not significant 
for pair~2, which seems consistent with a burst and a
comparatively long time delay; for pairs~1 and 3, however, 
the likelihood peak is at least a factor 10 higher than any 
value in the domain $\tau_{100}>10\,$yr.
Using Eq.~(\ref{t_delay}), this can be transformed into the
tentative limit
\begin{equation}
  B_{\rm rms}\lesssim6\times10^{-11}
  \left(\frac{l_{\rm c}}{1\,{\rm Mpc}}\right)^{-1/2}
  \left(\frac{D}{30\,{\rm Mpc}}\right)^{-1}\,{\rm G}
  \,.\label{limit}
\end{equation}
If confirmed, or possibly even improved by future data, this
constraint would be more stringent than existing 
limits~\cite{Kronberg} by more than an order of magnitude.

\end{itemize}

We also performed runs for $D=10\,$Mpc and $D=60\,$Mpc, as
well as for $n_B=4$ and for $l_{\rm c}=10\,$Mpc. For corresponding
values of the parameters in Eq.~(\ref{likelihood}), the
likelihood does not significantly depend on $n_B$ and $l_{\rm c}$.
For $D\lesssim60\,$Mpc the dependence on $D$ is also insignificant,
except possibly for pair~1 which favors
$D\lesssim30\,$Mpc within a factor $3-4$ in the likelihood.

Furthermore, we performed simulations for sources located
within an extended galactic halo, in which case the time delay is
dominated by the halo magnetic field.
We chose a halo extension of $100\,$kpc with a coherence
scale  $l_{\rm c}=30\,$kpc,  $n_B=4$, although the results
again are not sensitive to these parameters. The maxima 
of the likelihood do not significantly differ from those 
obtained previously for a dominant extra-galactic magnetic 
field. The main difference between these two limits is the 
predominance of a soft spectrum $\gamma\simeq2.0$ for pair~1, 
in the halo field case, even though this pair includes the 
$200\,$EeV event. This is demonstrated in Fig.~\ref{F8} and
can be attributed to the absence of frequent pion
production. Furthermore, if $D$ is now interpreted as the
minimum of
the source distance and the extension of the magnetized halo,
Eq.~(\ref{limit}) translates into the tentative constraint on the
halo field
\begin{equation}
  B_{\rm rms}\lesssim6\times10^{-7}
  \left(\frac{l_{\rm c}}{1\,{\rm kpc}}\right)^{-1/2}
  \left(\frac{D}{100\,{\rm kpc}}\right)^{-1}\,{\rm G}
  \,.\label{limit1}
\end{equation}

Finally, we mention that the results presented in this section
do not depend significantly on the lower boundary $E_0$ of the
energy range over which we compute the likelihood function
Eq.~(\ref{likelihood}) as long as it is not much lower than the lowest
event energy in the pairs, within the energy
resolution. Lowering $E_0$ tends to make the $\gamma-$dependence
of the likelihood function more significant, but neither a hard
nor a soft injection spectrum can be excluded for the sources of
the three AGASA pairs.

\section{Discussion and Conclusions}

Under our assumptions on the configuration of the 
LSMF, namely a power-law power spectrum
down to some cutoff scale $l_{\rm c}\gtrsim1\,$Mpc and
$l_{\rm c}\gtrsim1\,$kpc for time delays dominated by the
extra-galactic or the galactic halo field, respectively, and within 
the limited range probed for the average time delay $\tau_{100}$, our 
analysis suggests that all three AGASA pairs originated
in different sources! Pair~1 would originate from a burst with a 
short time delay, $\tau_{100}\sim1{\rm yr}$, i.e., a time delay 
inconsistent with that required in a cosmological GRB model. 
Pair~2 could originate from a burst, if the average time delay 
is sufficiently large, $\tau_{100}\gtrsim50{\rm yr}$, as advocated  
in the cosmological GRB model, although it marginally favors a 
continuously emitting source with a high energy output. Finally, 
pair~3 cannot originate from a burst, unless the time delay is 
extremely large, and probably too large when compared to the 
Faraday rotation bound; it has to originate within a 
continuously emitting source, with $T_{\rm S}\gtrsim100{\rm yr}$, 
and $N_0\gtrsim100$, in agreement with conventional sources 
such as powerful radio-galaxies, for instance.

Let us now address the possibility of bursting topological
defect sources. For example, certain classes of cosmic string
loops might collapse and release all of their energy in form of 
UHECRs within about one light crossing time
$T_{\rm S}\ll1\,$yr~\cite{BR},
for symmetry breaking scales near the grand unified scale and
typical total burst energies $\sim10^{51}\,$erg.
For an LSMF $\gtrsim10^{-11}\,$G, events above
$\simeq80\,$EeV are predicted to be most likely $\gamma$-rays,
whereas around $50\,$EeV an approximately equal amount of
protons is expected~\cite{SLC}. To take into
account the likely $\gamma-$ray nature of the higher energy
events in these models, a more accurate treatment should
include propagation and deflection of electromagnetic cascades
in the large-scale magnetic field. However, since for a given
energy the amount by which an electromagnetic cascade particle
and a nucleon is deflected and delayed is comparable, the 
topological defect scenario for the origin of UHECRs is not
inconsistent with our above results, notably as far as pair~1 is 
concerned. In this respect, we note that the muon
contents of the AGASA pairs air showers are not in
contradiction with interpreting the higher energy event as a
$\gamma$-ray~\cite{Hayashida1}. The most efficient test of the 
topological defect scenario clearly resides in the chemical
composition at energies $\gtrsim80{\rm EeV}$, and improved data
on UHECR composition could rule out this model in the future.

We do not wish to emphasize too strongly the above 
results, as quite a number of approximations had to be made, that
restrict the range of parameter space available to us, and, 
furthermore, the statistics of the AGASA observations are 
limited. The main objective of the present work was, rather, to 
introduce the numerical code developed, and to give an example 
of what could be achieved by future detectors. From our
experience, it seems reasonable to say that one or several clusters of
more than $\sim10$ particles per source, would allow to pin down 
most of the parameters, notably: the magnetic field strength 
$B_{\rm rms}$ and coherence length $l_{\rm c}$, the distance to 
the source $D$, the emission time scale $T_{\rm S}$, and the time 
delay $\tau_{100}$. Other parameters such as the 
differential energy index $\gamma$ and the index of the 
power spectrum of magnetic inhomogeneities $n_B$ (if any), are 
more difficult to grasp. In a separate paper, we present 
qualitative aspects of the angle-time-energy images of UHECR
sources~\cite{LSOS}; the qualitative features given there
would allow to considerably restrain the size of the parameter
space, and even give estimates of some of the above parameters. This
would constitute a very useful first approach to a full
likelihood evaluation.

In Ref.~\cite{LOS} it was shown that the $\gamma-$ray flux
above $10^{19}\,$eV contains information about the strength of 
the extra-galactic magnetic field. Here, we have shown that a
likelihood analysis of clusters of charged UHECRs from a common
extra-galactic source is also sensitive to  the strength and
the coherence scale  in a range $10^{-12}{\rm G}\lesssim B_{\rm 
rms}\lesssim10^{-9}{\rm G}$, with $l_{\rm c}\gtrsim1{\rm Mpc}$,
and in addition to a possible galactic halo field in the range
$10^{-8}\,{\rm G}\lesssim B_{\rm rms}\lesssim10^{-6}\,$G, with
$l_{\rm c}\gtrsim1\,$kpc. The deflection and delay of such UHECR
clusters can thus be used as a new tool to probe the LSMF in a
regime that is intermediate between field strengths accessible
to conventional methods such as measuring the Faraday rotation
of polarized light (see, e.g., Ref.~\cite{Kronberg}), and much
weaker fields that could be detected by observing
electromagnetic cascades in the TeV range~\cite{Plaga,WC}. The
UHECR pairs observed by AGASA already indicate a trend towards
comparatively small time delays, that can be translated into a
tentative upper limit on the magnetic field strength, roughly
an order of magnitude better than the Faraday rotation limit.

Future instruments in construction or in the proposal stage such
as the Japanese Telescope Array~\cite{Teshima},
the High Resolution Fly's Eye~\cite{Bird4},
and the Pierre Auger Project~\cite{Cronin}
will have the potential to test whether there is
significant clustering of UHECRs. The latter experiment, with
an angular resolution of a fraction of $1^\circ$ and an energy
resolution of $\simeq10\%$, should detect clusters of $20-50$
events if the clustering observed by AGASA is real. With
such statistics, it will be possible to answer the question on
the nature of the sources of such UHECR clusters, and, in turn,
to use these as ``candles'' to probe cosmic magnetic fields.

\section*{Acknowledgments}
We acknowledge P.~Biermann, A.~Dubey, C.~Graziani, J.~Quashnock,
and D.~Schramm for useful discussions. The Aspen Center for
Physics is thanked for hospitality and support. We are grateful to Sangjin 
Lee for providing us with suitable pion and pair production
tables from his thesis work. The Max-Planck
Institut f\"ur Physik, M\"unchen, Germany and the Institut 
d'Astrophysique de Paris, Paris,
France, are thanked for providing CPU time.
G.S. acknowledges financial support by the Deutsche Forschungs
Gemeinschaft under grant SFB 375 and by the Max-Planck Institut
f\"ur Physik.
This work was supported, in part,
by the DoE, NSF, and NASA at the University of Chicago, and
by the DoE and by NASA through grant NAG 5-2788 at Fermilab.


\end{document}